\documentclass{aastex}

\usepackage{emulateapj5}
\usepackage{aastexug}
\usepackage{amsmath}
\usepackage{graphicx}
\usepackage{subfigure}
\usepackage{apjfonts}

\def\gsim{\;\lower4pt\hbox{${\buildrel\displaystyle >\over\sim}$}\;}
\def\lsim{\;\lower4pt\hbox{${\buildrel\displaystyle <\over\sim}$}\;}
\def\grls{\;\lower4pt\hbox{${\buildrel\displaystyle >\over <}$}\;}

\begin{document}

\title{Global axisymmetric stability analysis for a composite
   system of two gravitationally coupled scale-free discs}
\shorttitle{Axisymmetric stability}

\shortauthors{Shen \& Lou}

\author{Yue Shen$^{1}$ and Yu-Qing Lou$^{1,2,3}$}

\affil{$^1$Physics Department, The Tsinghua Center for
Astrophysics, Tsinghua University, Beijing 100084, China\\
$^2$Department of Astronomy and Astrophysics, The University
of Chicago, 5640 S. Ellis Ave., Chicago, IL 60637 USA\\
$^3$National Astronomical Observatories, Chinese Academy of
Sciences, A20, Datun Road, Beijing 100012, China.}


   \begin{abstract}
   In a composite system of gravitationally coupled stellar
   and gaseous discs, we perform linear stability analysis for
   axisymmetric coplanar perturbations using the two-fluid formalism.
   The background stellar and gaseous discs are taken to be scale-free
   with all physical variables varying as powers of cylindrical radius
   $r$ with compatible exponents. The unstable modes set in as neutral
   modes or stationary perturbation configurations with angular
   frequency $\omega=0$. The axisymmetrically stable range is bounded
   by two marginal stability curves derived for stationary perturbation
   configurations. By the gravitational coupling between the stellar
   and the gaseous disc components, one only needs to consider the
   parameter regime of the stellar disc. There exist two unstable
   regimes in general: the collapse regime corresponding to large-scale
   perturbations and the ring-fragmentation regime corresponding to
   short-wavelength perturbations. The composite system will collapse
   if it rotates too slowly and will succumb to ring-fragmentation
   instabilities if it rotates sufficiently fast. The overall stable
   range for axisymmetric perturbations is determined by a necessary
   $D-$criterion involving of the effective Mach number squared
   $D_s^2$ (i.e., the square ratio of the stellar disc rotation speed
   to the stellar velocity dispersion but scaled by a numerical factor).
   Different mass ratio $\delta$ and sound speed ratio $\eta$ of the
   gaseous and stellar disc components will alter the overall stability.
   For spiral galaxies or circumnuclear discs, we further include the
   dynamical effect of a massive dark matter halo. As examples,
   astrophysical applications to disc galaxies, proto-stellar discs
   and circumnuclear discs are discussed.
   \end{abstract}

\section{Introduction}       
\label{sect:intro}
Axisymmetric instabilities in models of disc galaxies have been
investigated extensively in the last century (e.g., Safronov
1960; Toomre 1964; Binney \& Tremaine 1987; Bertin \& Lin 1996).
For a single disc of either gaseous or stellar content, Safronov
(1960) and Toomre (1964) originally introduced a dimensionless
$Q$ parameter to determine the local stability condition (i.e.
$Q>1$) against axisymmetric ring-like disturbances in the usual
Wentzel-Kramers-Brillouin-Jeffreys (WKBJ) or tight-winding
approximation. A more realistic model of a disc galaxy would
involve both gas and stars as well as an unseen massive dark
matter halo, all interacting among themselves through the mutual
gravitation.\footnote{Magnetic field and cosmic-ray gas component
are dynamically important on large scales (Fan \& Lou 1996; Lou
\& Fan 1998a, 2003; Lou \& Zou 2004 and references therein) in
the galactic gas disc of interstellar medium (ISM) but are not
considered here for simplicity.}
Many theoretical investigations have been conducted along this track
for a composite system of two coupled discs (Lin \& Shu 1966; Kato
1972; Jog \& Solomon 1984a, b; Bertin \& Romeo 1988; Romeo 1992;
Elmegreen 1995; Jog 1996; Lou \& Fan 1998b). In these earlier
treatments, either an approach of combined distribution function and
fluid or the formalism of two fluids have been adopted in a WKBJ modal
analysis. While these model results were initially derived in various
galactic contexts, they can be applied or adapted also to, with proper
qualifications, relevant self-gravitating disc systems including
accretion discs in general, circumnuclear discs, protostellar discs,
planetary discs and so forth.

The local WKBJ or tight-winding approximation has been proven to
be a powerful technique in analysing disc wave dynamics. Meanwhile,
theorists have long been keenly interested in a class of relatively
simple disc models referred to as scale-free discs (Mestel 1963;
Zang 1976; Lemos et al. 1991; Lynden-Bell \& Lemos 1993; Syer \&
Tremaine 1996; Evans \& Read 1998; Goodman \& Evans 1999; Shu et
al. 2000; Lou 2002; Lou \& Fan 2002; Lou \& Shen 2003; Shen \& Lou
2003; Lou \& Zou 2004; Lou \& Wu 2004). Scale-free discs, where all
pertinent physical variables (e.g., disc rotation speed, surface
mass density, angular speed etc.) scale as powers of cylindrical
radius $r$, have become one effective and simple vehicle to explore
disc dynamics. Perhaps, the most familiar case is the so-called
singular isothermal discs (SIDs) or Mestel discs with an isothermal
equation of state and flat rotation curves (Mestel 1963; Zang 1976;
Goodman \& Evans 1999; Shu et al. 2000; Lou 2002; Lou \& Shen 2003;
Lou \& Zou 2004). In contrast to the usual WKBJ approximation for
perturbations, perturbations in axisymmetric scale-free discs can
be treated, in some cases, globally and exactly without the local
restriction (i.e., valid only in the short-wavelength regime). It is
therefore possible to derive global properties of perturbations. Using
scale-free disc models, Lemos et al. (1991) and Syer \& Tremaine (1996)
both studied the axisymmetric stability problem for a single disc and
found that instabilities first set in as neutral modes or stationary
configurations with angular frequency $\omega=0$.

The main motivation of this paper is to examine the global
axisymmetric stability problem in a composite system of two
gravitationally coupled scale-free discs. As a more general
extension to the previous two-SID analysis (Lou \& Shen 2003;
Shen \& Lou 2003), we further consider a much broader class
of rotation curves as well as the equation of state. This
contribution gives an explicit proof that stationary
configurations ($\omega=0$) do mark the marginal stability
in the two-fluid system, a cogent supplement to our recent
investigation on stationary perturbation configurations
(Shen \& Lou 2004).

\section{Two-Fluid Formalism}
\label{sect:form}

As an expedient approximation, we treat both discs as razor-thin
discs and use either superscripts or subscripts $s$ and $g$ to
indicate physical variables with stellar and gaseous disc
associations, respectively. The large-scale coupling between
the two discs is primarily caused by the mutual gravitational
interaction. In the present formulation of large-scale
perturbations, we ignore non-ideal diffusive effects such as
viscosity, resistivity and thermal conduction, etc. It is then
straightforward to write down the basic coplanar fluid equation
set for the stellar disc in cylindrical coordinates
$(r,\ \theta,\ z)$ within the $z=0$ plane, namely
\begin{equation}
\frac{\partial \Sigma^{s}}{\partial t}
+\frac{1}{r}\frac{\partial }{\partial r} (r\Sigma^{s}u^{s})
+\frac{1}{r^{2}}\frac{\partial }{\partial \theta }
(\Sigma^{s}j^{s})=0\ ,
\end{equation}
\begin{equation}
\frac{\partial u^{s}}{\partial t}
+u^{s}\frac{\partial u^{s}}{\partial r}
+\frac{j^{s}}{r^{2}}\frac{\partial u^{s}}
{\partial \theta }-\frac{j^{s2}}{r^{3}}
=-\frac{1}{\Sigma^{s}}\frac{\partial \Pi^s}
{\partial r}-\frac{\partial \phi }{\partial r}\ ,
\end{equation}
\begin{equation}
\frac{\partial j^{s}}{\partial t}
+u^{s}\frac{\partial j^{s}}{\partial r}
+\frac{j^{s}}{r^{2}}\frac{\partial j^{s}}
{\partial \theta } =-\frac{1}{\Sigma^{s}}
\frac{\partial\Pi^s } {\partial \theta }
-\frac{\partial \phi }{\partial \theta }\ ,
\end{equation}
where $\Sigma^s$ is the surface mass density, $u^s$ is the radial
bulk flow velocity, $j^s$ is the specific angular momentum about
the rotation axis along $z-$direction, $\Pi^s$ is the vertically
integrated effective (two-dimensional) pressure due to stellar
velocity dispersion and $\phi$ is the {\it total} gravitational
potential. For the gaseous disc, we simply replace superscript or
subscript $s$ by $g$ in the above three equations. The coupling
of the two sets of fluid equations is due to the gravitational
potential through the Poisson integral
\begin{eqnarray}
\phi(r,\theta,t)=
\oint\!d\psi\!\!\int_0^{\infty}\!\!
\frac{-G\Sigma (r^{\prime},\psi ,t)r^{\prime }
dr^{\prime }}{[r^{\prime 2}+r^{2}-2rr^{\prime}
\cos (\psi -\theta )]^{1/2}}\ ,
\end{eqnarray}
where $\Sigma =\Sigma^{s}+\Sigma^{g}$ is the total
surface mass density of the composite disc system.

The barotropic equation of state assumes the relation between
two-dimensional pressure and surface mass density in the form of
\begin{equation}
\Pi=K\Sigma^{n},
\end{equation}
where coefficients $K\geq0$ and $n>0$ are constant. This directly
leads to the sound speed\footnote{In a stellar disc, the velocity
dispersion mimics the sound speed to some extent.  }
$a$ defined by
\begin{equation}
a^2=\frac{d\Pi_0}{d\Sigma_0}=n K\Sigma_0^{n-1},
\end{equation}
which scales as $\propto\Sigma_0^{n-1}$. The case of
$n=1$ corresponds to an isothermal sound speed $a$.

From the basic fluid equations for stellar and gaseous discs above
(the latter are not written out explicitly), we may derive the
axisymmetric equilibrium background properties (Shen \& Lou 2004).
We presume that in the equilibrium background of axisymmetry, both
rotation curves of the two discs scale as $\propto r^{-\beta}$ and
both surface mass densities of the two discs scale as $\propto
r^{-\alpha}$ where $\alpha$ and $\beta$ are two constant exponents
and the proportional coefficients are allowed to be different in
general. By assuming the same power-law indices for the stellar
and gaseous disc rotation curves (or equivalently the surface
mass densities), it is possible to consistently construct a global
axisymmetric background equilibrium for the composite disc system
that meets the requirement of radial force balance at all radii
[see equation (2) and the corresponding equation with the
superscript $s$ replaced by $g$] and satisfies the Poisson
integral (4) simultaneously.

The scale-free condition requires the following relationship
among $\alpha$, $\beta$ and $n$ (see Syer \& Tremaine 1996
and Shen \& Lou 2004), namely
\begin{equation}
\alpha=1+2\beta \ \hbox{ and }\  n=\frac{1+4\beta}{1+2\beta}\ .
\end{equation}
Once the rotation curve is specified, all other
physical variables are simultaneously determined.

With the knowledge of computing a gravitational potential
arising from an axisymmetric power-law surface mass
density for the background rotational equilibrium (Kalnajs
1971; Qian 1992; Syer \& Tremaine 1996), we can derive a
self-consistent axisymmetric background equilibrium
surface mass densities as
\begin{equation}
\begin{split}
\Sigma_0^s=\frac{A_s^2(D_s^2+1)} {2\pi G(2\beta{\cal
P}_0)r^{1+2\beta}(1+\delta)},\
\Sigma_0^g=\frac{A_g^2(D_g^2+1)\delta} {2\pi G(2\beta{\cal
P}_0)r^{1+2\beta}(1+\delta)}\ ,
\end{split}
\end{equation}
where the coefficient ${\cal P}_0$ as a function
of $\beta$ involves $\Gamma-$functions
\begin{equation}\label{P0}
{\cal P}_0\equiv
\frac{\Gamma(-\beta+1/2)\Gamma(\beta)}
{2\Gamma(-\beta+1)\Gamma(\beta+1/2)}\
\end{equation}
and parameter $\delta\equiv\Sigma_0^g/\Sigma_0^s$ is the ratio of
the surface mass density of the gaseous disc to that of the
stellar disc, $A$ and $D$ are two dimensionless parameters. We
note that the value of $2\beta{\cal P}_0$ falls within
$(0,\infty)$ for the prescribed range\footnote{The valid range of
$\beta\in(-1/4,1/2)$ is determined by (1) barotropic index $n>0$
for warm discs, (2) surface mass density exponent $\alpha<2$ so
that the central point mass does not diverge, and (3) this $\beta$
range is contained within a wider range of $\beta\in(-1/2,1/2)$
(for a cold disc system) when the computed force arising from the
background potential remains finite (Syer \& Tremaine 1996).} of
$\beta\in(-1/4,1/2)$ and is equal to 1 when $\beta\rightarrow 0$
(i.e., the case of two gravitationally coupled SIDs).

From the radial force balance of the background
equilibrium, there also exists a relation such that
\begin{equation}\label{ADeqn}
A_s^2(D_s^2+1)=A_g^2(D_g^2+1)\ ,
\end{equation}
where $\eta\equiv A_s^2/A_g^2=a_s^2/a_g^2$ is another
handy dimensionless parameter for sound speed ratio
squared. We note that $A$ is actually
the reduced\footnote{By the adjective ``reduced'', we refer to
the part of a physical variable after removing the power-law
radial dependence. For example, in the disc rotation speed
$v={\cal V}r^{-\beta}$, quantity ${\cal V}$ is referred to as
the reduced disc rotation speed.}
effective sound speed [scaled by a factor $(1+2\beta)^{1/2}$]
and the parameter $D\equiv {\cal V}/A$ where ${\cal V}$ is
the reduced disc rotation speed, is the effective Mach
number [see eqns (\ref{j0})$-$(\ref{soundspeed}) later].
Condition (\ref{ADeqn}) is very important in our analysis
because the rotations of the two discs are not independent
of each other but are dynamically coupled. It suffices to
examine the parameter regime in either stellar or gaseous
disc. In disc galaxies, the typical velocity dispersion in
a stellar disc exceeds the sound speed in a gaseous disc,
implying $\eta>1$ so that inequality $D_g^2>D_s^2$ holds.
Therefore, the physical requirement $D_s^2>0$ absolutely
guarantees $D_g^2>0$ and it suffices to consider the
stability problem in terms of $D_s^2>0$ together with
parameters $\delta$ and $\eta$ for different values [note
that $D_g^2=\eta(D_s^2+1)-1$]. With these explanations, we
shall express other equilibrium physical variables in terms
of parameters $A$ and $D$.

The specific angular momenta $j_0^s$ and $j_0^{g}$ about
the $z-$axis and the sound speeds $a_s$ and $a_g$ in the
two coupled equilibrium discs are expressed by
\begin{equation}\label{j0}
j_0^s=A_sD_sr^{1-\beta}\ , \qquad j_0^{g}=A_gD_gr^{1-\beta}\ ,
\end{equation}
\begin{equation}\label{soundspeed}
\begin{split}
&a_s^2=n K_s(\Sigma_0^s)^{n-1}=A_s^2/[(1+2\beta)r^{2\beta}]\ ,\\
&a_g^2=n K_g(\Sigma_0^g)^{n-1}=A_g^2/[(1+2\beta)r^{2\beta}]\ .
\end{split}
\end{equation}
The disc angular rotation speed $\Omega\equiv j_0/r^2$ and the
epicyclic frequency $\kappa\equiv[(2\Omega/r)d(r^2\Omega)/dr]^{1/2}$
are similarly expressed in terms of $A$ and $D$ as
\begin{equation}
\begin{split}
&\Omega_s=A_sD_sr^{-1-\beta}\ , \qquad
\kappa_s=[2(1-\beta)]^{1/2}\Omega_s\ ,\\
&\Omega_g=A_gD_gr^{-1-\beta}\ , \qquad
\kappa_g=[2(1-\beta)]^{1/2}\Omega_g\ ,\\
\end{split}
\end{equation}
with $dj_0/dr=r\kappa^2/(2\Omega)$
to simplify later derivations.

\subsection{Linear Perturbation Equations}
\label{subsect:linear}

For a composite system of two gravitationally coupled discs
in rotational equilibrium with axisymmetry, we introduce
small coplanar perturbations denoted by subscript $1$ along
relevant physical variables. The corresponding linearized
perturbation equations can be derived from the basic nonlinear
equations (1)$-$(4) as
\begin{equation}\label{perturb0}
\begin{split}
&\frac{\partial \Sigma_1^{s}}{\partial t}
+\frac{1}{r}\frac{\partial }{\partial r}
(r\Sigma_0^{s}u_1^{s})+\Omega_s\frac{\partial
\Sigma_1^s}{\partial\theta}
+\frac{\Sigma_0^{s}}{r^2}\frac{\partial j_1^{s}}
{\partial \theta}=0\ ,\\
&\frac{\partial u_1^{s}}{\partial t}
+\Omega_s\frac{\partial u_1^{s}}{\partial \theta}
-2\Omega_s\frac{j_1^{s}}{r}=-\frac{\partial}{\partial r}
\bigg(a_s^2\frac{\Sigma_1^{s}}{\Sigma_0^{s}}+\phi_1\bigg)\ ,\\
&\frac{\partial j_1^{s}}{\partial t}
+\frac{r\kappa_s^2}{2\Omega_s}u_1^{s}
+\Omega_s\frac{\partial j_1^{s}}{\partial\theta}
=-\frac{\partial}{\partial\theta}
\bigg(a_s^2\frac{\Sigma_1^{s}}{\Sigma_0^{s}}+\phi_1\bigg)\
\end{split}
\end{equation}
for the stellar disc as well as their counterparts for
the gaseous discs, together with the Poisson integral
\begin{equation}\label{perturbV0}
\phi_1(r,\theta,t)=
\oint\!d\psi\!\!\int_0^{\infty}\frac{-G(\Sigma_1^{s}
+\Sigma_1^{g})r^{\prime}dr^{\prime}}
{\left[ r^{\prime 2}+r^{2}-2rr^{\prime}
\cos(\psi-\theta )\right]^{1/2}}\
\end{equation}
relating the total gravitational potential perturbation
$\phi_1$ and the total surface mass density perturbation
$\Sigma_1\equiv\Sigma_1^{s}+\Sigma_1^{g}$.

Given a Fourier periodic component in the form of
exp$[i(\omega t-m\theta)]$ for small perturbations
in general,
we write for coplanar perturbations in the stellar disc
\begin{equation}\label{fouriers}
\begin{split}
&\Sigma_1^s=\mu^s(r)\hbox{exp}[i(\omega t-m\theta)]\ ,\\
&u_1^s=U^s(r)\hbox{exp}[i(\omega t-m\theta)]\ ,\\
&j_1^s=J^s(r)\hbox{exp}[i(\omega t-m\theta)]\ ,\\
\end{split}
\end{equation}
as well as their counterparts in the gaseous disc, together
with the total gravitational potential perturbation
\begin{equation}\label{fourierV}
\phi_1=V(r)\hbox{exp}[i(\omega t-m\theta)]\ ,
\end{equation}
where integer $m$ is taken to be non-negative. For
axisymmetric $m=0$ perturbations, we introduce
Fourier decompositions in equations (\ref{perturb0})
$-$(\ref{perturbV0}) for the stellar disc to deive
\begin{equation}\label{perturbs1}
\begin{split}
&i\omega\mu^s+\frac{1}{r}\frac{d}{dr}(r\Sigma_0^sU^s)
=0\ ,\\
&i\omega U^s-2\Omega_s\frac{J^s}{r}
=-\frac{d}{dr}\bigg(a_s^2\frac{\mu^s}{\Sigma_0^s}+V\bigg)\ ,\\
&i\omega J^s+\frac{r\kappa_s^2}{2\Omega_s}U^s =0\ .
\end{split}
\end{equation}
We do the same in parallel for the gaseous disc to derive
\begin{equation}\label{perturbg1}
\begin{split}
&i\omega \mu^g+\frac{1}{r}\frac{d}{dr}(r\Sigma_0^gU^g)
=0\ ,\\
&i\omega U^g-2\Omega_g\frac{J^g}{r}
=-\frac{d}{dr}\bigg(a_g^2\frac{\mu^g}{\Sigma_0^g}+V\bigg)\ ,\\
&i\omega J^g+\frac{r\kappa_g^2}{2\Omega_g}U^g =0\ .
\end{split}
\end{equation}
For the total gravitational potential
perturbation, we simply have
\begin{equation}\label{perturbV1}
V(r)=\oint\!d\psi\!\!\int_0^{\infty}\frac{-G(\mu^{s} +\mu^{g})\cos
(m\psi)r^{\prime}dr^{\prime }}{(r^{\prime 2}+r^{2} -2rr^{\prime
}\cos \psi)^{1/2}}\ .
\end{equation}
Equations (\ref{perturbs1})$-$(\ref{perturbV1}) are
the basic coplanar perturbation equations used for
our axisymmetric stability analysis.

\subsection{Axisymmetric Stability Analysis}
\label{subsect:axisy}

For axisymmetric stability analysis with radial oscillations,
we choose the Kalnajs potential-density pairs below because
perturbations can be generally expanded in terms of such
complete basis functions (Kalnajs 1971; Binney \& Tremaine
1987; Lemos et al. 1991; Lou \& Shen 2003). Specifically, we
take
\begin{equation}\label{spirPerturb}
\begin{split}
&\mu^s=\sigma^sr^{-3/2}\exp(i\xi\ln r)\ , \qquad
\mu^g=\sigma^gr^{-3/2}\exp(i\xi\ln r)\ ,\\
&V=-2\pi Gr(\mu^s+\mu^g){\cal N}_m(\xi)\ ,
\end{split}
\end{equation}
where $\xi$ is a `wavenumber' characterizing the radial
variation scale, $\sigma^s$ and $\sigma^g$ are two small
real coefficients and the parameter function
\begin{equation}
{\cal N}_m(\xi)
=\frac{\Gamma(m/2+i\xi/2+1/4)\Gamma(m/2-i\xi/2+1/4)}
{2\Gamma(m/2+i\xi/2+3/4)\Gamma(m/2-i\xi/2+3/4)}
\end{equation}
is the Kalnajs function (Kalnajs 1971) that involves
$\Gamma-$functions of complex arguments. Note that
${\cal N}_m$ is even in $\xi$. It then suffices to
consider only $\xi\ge 0$.

Using the first mass conservations in equations
(\ref{perturbs1}) and (\ref{perturbg1}) respectively,
we infer $U\propto i\omega r^{1/2+2\beta+i\xi}$ (see
also Lou \& Zou 2004). By potential-density pair
(\ref{spirPerturb}), equations (\ref{perturbs1}) and
(\ref{perturbg1}) reduce to
\begin{equation}\label{dr1}
\begin{split}
&(\omega^2-H_1)U^s=-G_2U^g\ ,\\
&(\omega^2-H_2)U^g=-G_1U^s\
\end{split}
\end{equation}
in the limit of $\omega\rightarrow 0$, where parameter
functions $H_1$, $H_2$, $G_1$ and $G_2$ are explicitly
defined by
\begin{equation}
\begin{split}
&H_1\equiv\kappa_s^2+\bigg(\frac{a_s^2}{r}
-2\pi G{\cal N}_0\Sigma_0^s\bigg)\frac{(\xi^2+1/4)}{r}\ ,\\
&H_2\equiv\kappa_g^2+\bigg(\frac{a_g^2}{r}
-2\pi G{\cal N}_0\Sigma_0^g\bigg)\frac{(\xi^2+1/4)}{r}\ ,\\
&G_1\equiv 2\pi G{\cal N}_0\Sigma_0^s\frac{(\xi^2+1/4)}{r}>0\ ,\\
&G_2\equiv 2\pi G{\cal N}_0\Sigma_0^g\frac{(\xi^2+1/4)}{r}>0\ .
\end{split}
\end{equation}
The axisymmetric dispersion relation in the composite
system follows from equation (\ref{dr1})
\begin{equation}\label{quardomega2}
\omega^4-(H_1+H_2)\omega^2+(H_1H_2-G_1G_2)=0\
\end{equation}
in the limit of $\omega\rightarrow 0$, which is identical
in form with earlier results obtained in the WKBJ regime (Jog
\& Solomon 1984a; Shen \& Lou 2003). It would be of interest
to note that the conditions for the stellar disc and the
gaseous disc to be separately stable are $H_1>0$ and $H_2>0$,
respectively. It is also reminded here that in the familiar
WKBJ regime, the dispersion relation in a single disc is
\begin{equation}\label{drWKBJ}
\omega^2=\kappa^2+k^2a^2-2\pi G|k|\Sigma_0\ ,
\end{equation}
where $k$ is the radial wavenumber, whereas in the present
global analysis as applied to a single disc, we have in
the limit of $\omega\rightarrow 0$
\begin{equation}\label{drglobal}
\omega^2=\kappa^2+\frac{(\xi^2+1/4)}{r^2}a_s^2
-2\pi G\frac{(\xi^2+1/4){\cal N}_0}{r}\Sigma_0\ .
\end{equation}
If we replace ${\cal N}_0(\xi)$ approximately by
$(\xi^2+1/4)^{-1/2}$ in the asymptotic regime of $\xi\gg 1$, we
readily identify the correspondence between the effective
wavenumber $(\xi^2+1/4)^{1/2}$ and $|k|r$ in the WKBJ limit of
$|k|r\gg 1$ by directly comparing dispersion relations
(\ref{drWKBJ}) and (\ref{drglobal}). Physically, dispersion
relation (\ref{drglobal}) is more generally applicable beyond the
WKBJ regime and is globally accurate only in the limit of
$\omega\rightarrow 0$ (Shu et al. 2000).

Getting back to the composite system, there are two real
roots\footnote{One can show that the determinant of equation
(\ref{quardomega2}) $\Delta\equiv (H_1-H_2)^2+4G_1G_2>0$ is
always true. }
of $\omega^2$ of equation (\ref{quardomega2})
\begin{equation}
\omega_{\pm}^2=\frac{1}{2}\{(H_1+H_2)
\pm[(H_1+H_2)^2-4(H_1H_2-G_1G_2)]^{1/2}\}\ ,
\end{equation}
with $\omega_{+}^2$ root being always positive\footnote{If
$H_1+H_2\ge 0$, then $\omega_{+}^2>0$; otherwise if $H_1+H_2< 0$,
then at least one of $H_1$ and $H_2$ is negative. It therefore
follows that $H_1H_2-G_1G_2<0$ and hence $\omega_{+}^2>0$. }
(Shen \& Lou 2003). For the purpose of axisymmetric stability
analysis, we only need to examine $\omega_{-}^2$ root, namely
\begin{equation}\label{w2minus}
\omega_{-}^2=\frac{1}{2}\{(H_1+H_2)
-[(H_1+H_2)^2-4(H_1H_2-G_1G_2)]^{1/2}\}\ .
\end{equation}
As the right-hand side of the above equation (\ref{w2minus}) is
always real, axisymmetric instabilities set in as stationary
perturbation configurations with $\omega_{-}^2=0$ and leads to
the marginal stability condition
\begin{equation}\label{MarStaCon}
H_1H_2=G_1G_2\
\end{equation}
that requires inequality $H_1+H_2\ge 0$. This inequality can be
shown in a straightforward manner to be automatically satisfied
if equation (\ref{MarStaCon}) holds true. Let us first write
\begin{equation}
\begin{split}
H_1=F_1-G_1\ \qquad\hbox{ and }\qquad\  H_2=F_2-G_2\ ,
\end{split}
\end{equation}
where $F_1$ and $F_2$ are explicitly defined by
\begin{equation}
\begin{split}
F_1\equiv\kappa_s^2+\frac{(\xi^2+1/4)}{r^2}a_s^2>0\ ,\
F_2\equiv\kappa_g^2+\frac{(\xi^2+1/4)}{r^2}a_g^2>0\ .
\end{split}
\end{equation}
It then follows from condition (\ref{MarStaCon})
of $H_1H_2=G_1G_2$ that
\begin{equation}
F_1F_2=F_1G_2+F_2G_1\ .
\end{equation}
As $F_1$, $F_2$, $G_1$ and $G_2$ are all
positive, we immediately conclude that
\begin{equation}
F_1-G_1>0 \qquad\hbox{ and }\qquad F_2-G_2>0\ ,
\end{equation}
which finally leads to
\begin{equation}
H_1+H_2>0\ .
\end{equation}
The composite disc system becomes inevitably unstable
for $H_1+H_2<0$ because of $\omega_{-}^2<0$. Else if
$H_1+H_2\ge 0$ but with $H_1H_2-G_1G_2<0$, we again have
$\omega_{-}^2<0$ for instabilities. Only when $H_1+H_2\ge 0$
and $H_1H_2-G_1G_2>0$ at the same time can the composite disc
system be stable against axisymmetric coplanar perturbations.
This is an important necessary stability criterion for
a composite system of two gravitationally coupled discs.
In other words, once one disc is unstable by itself (i.e.,
$H_1<0$ or $H_2<0$ or both), the two-disc system must be
unstable; even if the two discs are both separately stable,
the composite disc system can still become unstable (i.e.,
$H_1>0$ and $H_2>0$ but $H_1H_2-G_1G_2<0$).

By inserting expressions of $H_1$, $H_2$, $G_1$ and $G_2$ into
the marginal stability condition (\ref{MarStaCon}) together with
requirements of the background rotational equilibrium, we readily
derive a quadratic equation in terms of $y\equiv D_s^2$, namely
\begin{equation}\label{quadratic}
C_2y^2+C_1y+C_0=0\ ,
\end{equation}
where the coefficients are explicitly defined by
\begin{equation}\label{coef}
\begin{split}
C_2\equiv &{\cal B}_0{\cal H}_0\eta\ ,\\
C_1\equiv &\bigg[({\cal B}_0-{\cal A}_0){\cal H}_0 +\frac{({\cal
A}_0+{\cal B}_0)({\cal H}_0-{\cal B}_0)} {(1+\delta
)}\bigg]\eta\\
&-\frac{({\cal A}_0+{\cal B}_0)
({\cal H}_0+{\cal B}_0\delta)}{(1+\delta )}\ ,\\
C_0\equiv &\bigg[-{\cal A}_0{\cal H}_0+\frac{({\cal A}_0 +{\cal
B}_0)({\cal H}_0-{\cal B}_0)}{(1+\delta )}\bigg]\eta\\
&+({\cal A}_0+{\cal B}_0)^2-\frac{({\cal A}_0+{\cal B}_0) ({\cal
H}_0+{\cal B}_0\delta)}{(1+\delta )}\ ,
\end{split}
\end{equation}
and
\begin{equation}
\begin{split}
&{\cal A}_0(\xi)      \equiv \xi^2+1/4\ ,\\
&{\cal B}_0(\beta)    \equiv (1+2\beta)(-2+2\beta)\ ,\\
&{\cal C}(\beta)      \equiv (1+2\beta)/(2\beta{\cal P}_0)\ ,\\
&{\cal H}_0(\beta,\xi)\equiv {\cal C}{\cal N}_0{\cal A}_0+{\cal B}_0\ .
\end{split}
\end{equation}
This quadratic equation (\ref{quadratic}) of $y\equiv D_s^2$ can
be readily proven to always have two real solutions (Shen \& Lou
2004). Only the positive portion of $D_s^2$ solutions can be
regarded as physically acceptable. Typically, there exist two
different regimes bounded by the marginal $D_s^2$ stability curves
that are unstable against axisymmetric perturbations, that is, the
collapse regime for long-wavelength perturbations and the
ring-fragmentation regime for short-wavelength perturbations. In
contrast to the short-wavelength WKBJ approximation, the collapse
regime is novel and exact. Systems with too fast a rotation
parameter $D_s^2$ will fall into the ring-fragmentation regime
(Safronov 1960; Toomre 1964; Syer \& Tremaine 1996; Lou \& Fan
1998a, b; Shu et al. 2000; Lou 2002; Lou \& Shen 2003, 2004; Shen
\& Lou 2003), while those with too slow a $D_s^2$ parameter will
fall into the collapse regime. Shown in Fig. 1 is an example of
illustration with $\beta=1/4$, $\eta=1$ and an unconstrained
$\delta$. We note that cases with $\eta=1$ are essentially the
same as those of a single disc (Shen \& Lou 2004). The boundaries
of the two regimes of instabilities shown in Fig. 1 vary with
different parameters $\eta>1$ and $\delta$ for chosen values of
$\beta$ that describes the entire scale-free radial profile of a
composite disc system (i.e., disc rotation curves, surface mass
densities and barotropic equation of state). Qualitatively, the
increase of either $\eta$ and $\delta$ will aggravate the
ring-fragmentation instability while suppress the large-scale
collapse instability (Shen \& Lou 2003). While this can be
directly seen from equation (\ref{quadratic}) in the relevant
parameter regimes, it can also be understood physically in terms
of the dynamical coupling between the two discs through condition
(\ref{ADeqn}) constrained by scale-free conditions. For a larger
$\eta$, the reduced gas disc rotation speed ${\cal V}_g$ will
exceed the reduced stellar disc rotation speed ${\cal V}_s$ by a
larger margin and this tend to prevent an overall collapse of a
composite disc system. In other words, a gaseous disc component
with a relatively lower sound speed seems to prevent a collapse
(Shen \& Lou 2003). More details and examples can be found in Shen
\& Lou (2004).

It is well known that a $Q$ parameter can be defined to determine
local axisymmetric stability of a single-disc system (Safronov
1960; Toomre 1964; Binney \& Tremaine 1987). For a composite disc
system, it has been attempted to introduce an effective
$Q_{\hbox{eff}}$ parameter (Elmegreen 1995; Jog 1996; Lou \& Fan
1998b). As a result of straightforward numerical computations, we
have recently introduced a powerful $D-$criterion for axisymmetric
stability of a composite system of two coupled SIDs (Shen \& Lou
2003). It is natural to further generalize this $D-$criterion for
axisymmetric stability of a composite system of two coupled
barotropic discs by straightforward numerical computations as
shown in Fig. 1. In the present case, it is more practical and
simple to use $D_s^2$ parameter.

\vglue 0.2cm \figurenum{1} \centerline{
\includegraphics[scale=0.4]{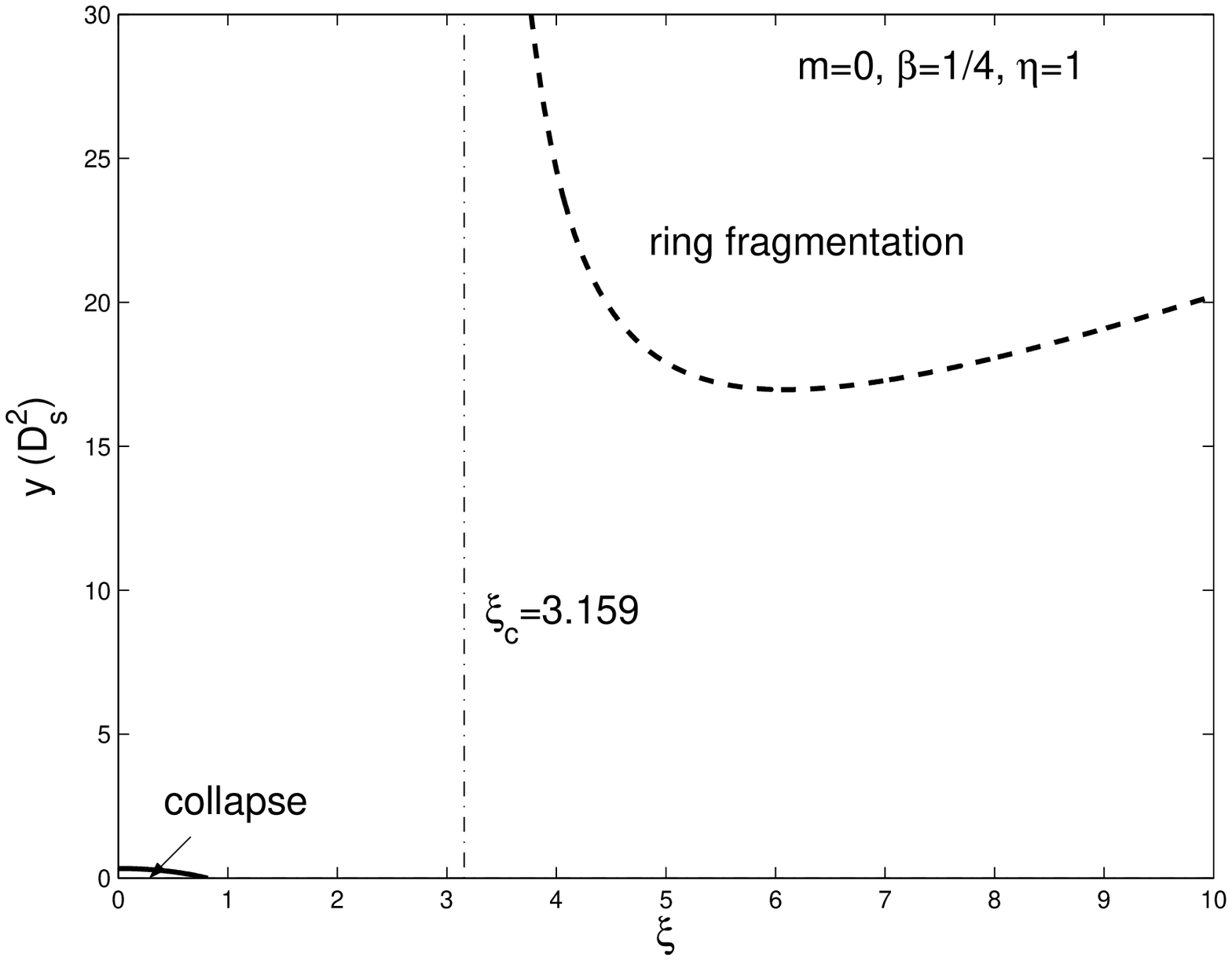}}
\figcaption{One illustrating example for the two unstable regimes
with $m=0$, $\beta=1/4$ and $\eta=1$. The collapse regime is at
the lower-left corner, while the ring-fragmentation regime is at
the upper-right corner. In this special case of $\eta=1$,
parameter $\delta$ can be arbitrary as can be seen from equation
(\ref{coef}), namely, when $\eta=1$ the coefficients $C_2$, $C_1$
and $C_0$ turn out to be independent of $\delta$. The vertical
dash-dotted line is the location of $\xi_c$ where ${\cal H}_0=0$
and $C_2$ vanishes. The solid line and the dashed line bound the
collapse regime and the ring fragmentation regime, respectively.
Only when $D_s^2$ falls within the range between the top of the
collapse regime and the bottom of the ring fragmentation regime
can a composite disc system become stable against all axisymmetric
coplanar perturbations. } \vglue 1cm

\subsection{Partial Disc Systems and Applications to Disc Galaxies}
\label{subsect:partial}

From observations of more or less flat rotation curves of most
disc galaxies, massive dark matter halos have been inferred to
exist ubiquitously as long as the Newtonian gravity remains
valid on galactic scales. If we naively attempt to relate
theoretical results obtained in Section 2.2 to a typical disc
galaxy, we may take the simple isothermal equation of state
as an example of illustration. The relevant parameters for a
composite SID system are then chosen as $\beta=0$ and
$a_s=50\hbox{ km s}^{-1}$,
${\cal V}_s=220\hbox{ km s}^{-1}$, $\delta=0.1$ and $\eta=50$.
Unfortunately, such a composite system of two coupled SIDs is
inevitably unstable against ring-fragmentation perturbations
because of a $D_s^2\simeq 20$, far exceeding the maximum value
necessary for being stable against ring fragmentations. This
dilemma can be resolved by attributing to an additional
gravitational potential associated with an unseen massive dark
matter halo. We refer to a composite disc system in association
with an axisymmetric dark matter halo as a composite system of
{\it partial discs} (e.g. Syer \& Tremaine 1996; Shu et al. 2000;
Lou 2002; Lou \& Fan 2002; Shen \& Lou 2003, 2004). In a simple
treatment, the dynamical effect of a dark matter halo is modeled
as only to contribute an axisymmetric gravitational potential
$\Phi$ in the background rotational equilibrium but not to
respond to coplanar perturbations in the composite disc system.
With $\phi_{T}\equiv \Phi+\phi$, we conveniently introduce
a dimensionless parameter ${\cal F}\equiv\phi/\phi_{T}$ as the
ratio of the potential arising from the composite disc system to
that of the entire system including the presumed axisymmetric
dark matter halo. The full-disc system
corresponds to ${\cal F}=1$ (i.e. $\Phi=0$) and the partial-disc
system corresponds to $0<{\cal F}<1$ (i.e., $\Phi\neq 0$). For
a composite system of two gravitationally coupled partial discs,
we follow the same procedure of analysing coplanar perturbations
in full discs to derive a similar quadratic equation of $D_s^2$
as the stationary dispersion relation. In a nutshell, we can
simply replace all ${\cal N}_0(\xi)$ in our theoretical results
by ${\cal F}{\cal N}_0(\xi)$ to accomplish this generalization
or extension. The introduction of the ratio parameter ${\cal F}$
will significantly reduce both the ring-fragmentation regime and
the collapse regime, as already can be seen from a comparative
study of the WKBJ and global approaches (Shen \& Lou 2003).

For the purpose of illustrating the stabilizing effect of a
partial-disc system, we simply take ${\cal F}=0.1$ with other
parameters used earlier in this section. The minimum value of
$D_s^2$ for unstable ring fragmentations now becomes $\sim 650$,
far beyond the actual value of $D_s^2\simeq 20$ in a disc galaxy.
Meanwhile, the collapse regime completely disappears. Therefore,
a typical composite system of two coupled  partial discd is
fairly stable against axisymmetric coplanar perturbations.

\section{Discussion and Summary}
\label{sect:summary}

The main thrust of this investigation is to model linear
coplanar perturbations of axisymmetry ($m=0$) in a composite
system of two-fluid scale-free discs with one intended for a
stellar disc and the other intended for a gaseous disc. The
two discs are dynamically coupled through the mutual
gravitational interaction. In order to include the dynamical
effect of a massive dark matter halo with axisymmetry, we
further describe a composite system of two coupled partial
discs (e.g. Syer \& Tremaine 1996; Shu et al. 2000; Lou 2002;
Lou \& Shen 2003; Lou \& Zou 2004; Lou \& Wu 2004). In a
global perturbation analysis, we show that axisymmetric
instabilities set in as stationary perturbation
configurations with $\omega=0$. The marginal $D_s^2$
stability curves (characterized by the stationary
configurations) delineate two different unstable regimes,
namely, the collapse regime for large-scale perturbations
and the ring-fragmentation regime for short-wavelength
perturbations. Apparently, the composite disc system becomes
less stable than a single-disc system and can be unstable
with the two discs being stable separately (Lou \& Fan 1998b).
In our analysis, stationary perturbation configurations turn
out to be more than just an alternative equilibrium state,
especially in view of the stability properties.

The basic results of this paper are generally applicable to
self-gravitating disc systems with or without axisymmetric dark
matter halos. The two-fluid treatment contains more realistic
elements than a single-disc formulation in the context of disc
galaxies. In addition to astrophysical applications to disc
galaxies, the studies presented here can be valuable for
exploring the dynamical evolution of protostellar discs and
circumnuclear discs.

In the context of a proto-stellar disc, it is the usual case to ignore
the self-gravity effect. By considering the self-gravity of a composite
disc system, our analysis indicates several qualitative yet interesting
results. For example, if the initial disc system rotates sufficiently
fast, the ring fragmentation (see the upper-right part of Fig. 1) can
occur at relatively small radial scales. By further non-axisymmetric
fragmentations, these condensed rings of materials may eventually become
birthplaces of planets. On the other hand, if the initial disc system
rotates sufficiently slow, then gravitational collapse can be induced by
perturbations of relatively large radial scales (see the lower left corner
of Fig. 1). Once such a perturbation develops in the background equilibrium
disc, it grows rapidly and destabilizes the disc. Subsequently, the system
undergoes global Jeans collapse to form a central young stellar object.
Finally, if the initial disc system rotates in a regime stable against all
axisymmetric perturbations (see Fig. 1), there might be two possibilities:
(1) the composite disc system might become unstable caused by
non-axisymmetric perturbations (not analyzed here) and
(2) the disc rotation may be gradually slowed down by some braking
mechanisms (e.g. magnetic field not included here and outflows or
winds) and the disc eventually succumbs to a central collapse induced
by large-scale perturbations.

Likewise, in the context of a circumnuclear disc around the center
of a galaxy, we can readily conceive similar physical processes in
parallel. One important distinction is that a dark-matter halo
should play an important dynamical role so that a formulation of a
partial composite disc system would be more appropriate. Here, the
ring fragmentation can be induced by relatively small-scale
perturbations in a disc system of sufficiently fast rotation. Such a
ring of relatively dense materials around the galactic center would
be a natural birthplace for circumnuclear starburst activities (e.g.
Lou et al. 2001). Depending on the evolution history of a circumnuclear
disc system, it may be stable initially and gradually lose angular
momentum by generating and damping spiral magnetohydrodynamic (MHD)
density waves (Lou et al. 2001). When the disc rotation becomes
sufficiently slow, Jeans collapse induced by large-scale perturbations
can set in to form a bulge or a supermassive black hole.

In summary, our global analysis shows the possible presence of an
evolution stage for a composite disc system against all axisymmetric
coplanar perturbations. More importantly, we reveal the parameter
regime of ring fragmentation and the parameter regime of large-scale
collapse. Astrophysical applications are discussed in the contexts
of disc galaxies, proto-stellar discs and circumnuclear disks.

\section*{Acknowledgments}
This research has been supported in part by the ASCI Center for
Astrophysical Thermonuclear Flashes at the University of Chicago
under Department of Energy contract B341495, by the Special Funds
for Major State Basic Science Research Projects of China, by the
Tsinghua Center for Astrophysics, by the Collaborative Research
Fund from the NSF of China (NSFC) for Young Outstanding Overseas
Chinese Scholars (NSFC 10028306) at the National Astronomical
Observatory, Chinese Academy of Sciences, by NSFC grant 10373009
at the Tsinghua University, and by the Yangtze Endowment from the
Ministry of Education through the Tsinghua University. Affiliated
institutions of Y.Q.L. share this contribution.

\end{document}